\documentclass[pdflatex,sn-vancouver-ay]{sn-jnl}
\usepackage{graphicx}
\usepackage{multirow}
\usepackage{xcolor}
\usepackage{textcomp}
\usepackage{amsmath}
\usepackage{longtable}
\usepackage{geometry}
\usepackage{array}

\title{\textbf{High hopes for \textit{Deep Medicine}? AI, economics, and the future of care}}

\author[1]{Robert Sparrow}

\author[2]{Joshua Hatherley}

\affil[1]{School of Philosophical, Historical, and International Studies, Monash University, Australia}

\affil[2]{Center for the Philosophy of AI, University of Copenhagen, Denmark}

\abstract{In the much‐celebrated book \textit{Deep Medicine}, Eric Topol argues that the development of artificial intelligence for health care will lead to a dramatic shift in the culture and practice of medicine. In the next several decades, he suggests, AI will become sophisticated enough that many of the everyday tasks of physicians could be delegated to it. Topol is perhaps the most articulate advocate of the benefits of AI in medicine, but he is hardly alone in spruiking its potential to allow physicians to dedicate more of their time and attention to providing empathetic care for their patients in the future. Unfortunately, several factors suggest a radically different picture for the future of health care. Far from facilitating a return to a time of closer doctor‐patient relationships, the use of medical AI seems likely to further erode therapeutic relationships and threaten professional and patient satisfaction.

\bigskip

This is a pre-print of: Sparrow, Robert and Joshua Hatherley. 2020. High hopes for "Deep Medicine"? AI, economics, and the future of care \textit{Hastings Center Report} 50(1): 14-17. \href{https://doi.org/10.1002/hast.1079}{10.1002/hast.1079}}

\begin{document}

\maketitle

\bigskip

In \textit{Deep Medicine}, Eric Topol \citeyearpar{topol2019deep} argues that the development of artificial intelligence (AI) for healthcare will lead to a dramatic shift in the culture and practice of medicine.  In the next several decades, he suggests, AI will become sophisticated enough for us to delegate many of the everyday tasks of physicians to it. According to Topol,

\begin{quote}
    The promise of artificial intelligence in medicine is to provide composite, panoramic views of individuals' medical data; to improve decision-making; to avoid errors such as misdiagnosis and unnecessary procedures; to help in the ordering and interpretation of appropriate tests; and to recommend treatment \citep[9]{topol2019deep}.\footnote{Later in the book he suggests that robots featuring AI might even perform some surgery \citep[161-162]{topol2019deep}. See also \cite{darzi2018better,liu2018time, verghese2018computer, wachter2017digital}}.
\end{quote} 

However, rather than replacing physicians, Topol suggests, AI could function alongside of them in order to allow them to devote more of their time to face-to-face patient care. Thus:

\begin{quote}
    The greatest opportunity offered by AI is not reducing errors or workloads, or even curing cancer: it is the opportunity restore the precious and time-honoured connection and trust — the human touch — between patients and doctors. Not only would we have more time to come together, enabling far deeper communication and compassion, but we would also be able to revamp how we select and train doctors… Eventually, doctors will adopt AI and algorithms as their work partners \citep[18]{topol2019deep}.
\end{quote}

Topol is perhaps the most articulate advocate of the benefits of AI in medicine, but he is hardly alone in spruiking its potential to allow physicians to dedicate more of their time and attention to providing empathetic care for their patients in the future \citep{israni2019humanizing, mesko2018will}.  Unfortunately, these high hopes for AI-enhanced medicine fail to appreciate a number of factors that, we believe, suggest a radically different picture for the future of healthcare. Far from facilitating a return to “the golden age of doctoring” \citep{mckinlay2002end},  the use of medical AI seems likely to further erode therapeutic relationships and threaten professional and patient satisfaction.

The fundamental problem with Topol’s optimistic vision for the future of medicine after AI is that substitutes fantasies about what AI might make possible for a realistic account of what it is likely to bring about.\footnote{There is more than a whiff of what \cite{broussard2018artificial} calls “technochauvinism” – the belief that smarter, faster, and cleaner technology is the solution to every problem – in Topol’s book, although it also includes passages that acknowledge that the impact of AI on medicine might be less positive than most of the text pretends, as we discuss further below.}  In particular, like many pundits who focus on technology rather than society when they think about the future, Topol neglects the role of economic and institutional considerations in determining how AI is likely to be used. 

The economics of healthcare, especially where it is provided in a for-profit context, will dictate that any time savings made possible by a reduction in the administrative burdens on physicians in the course of patient consultations will be used to move more patients through the system rather than to allow practitioners to spend more time talking with, and caring for, their patients. Even in the public sector, the institutional drive to cost savings and efficiency  prompted by concerns about the rising costs of healthcare \citep{diefenbach2009new}, as well as concerns about social justice in access to healthcare, are likely to mean that AI is likely to be used to improve access to healthcare, by increasing the number of people that a given service can treat per day,  rather than to increase the amount of time spent with each patient. 

Another powerful institutional dynamic may be expected to parallel and reinforce this economic imperative. Organizations tend to concentrate — one might even say fixate — on things that they can measure rather than the more subtle and intangible aspects of their operations \citep{blythe2018just}. Time per patient (or per procedure) is easily measured and optimized, whereas “care” is subtle and hard to measure. For this reason alone, there will be a tendency for institutions to use AI to treat more patients rather than devote more time to each patient.

Topol is conscious that institutions might adopt AI in ways that exacerbate rather than mitigate the dynamics that currently work to prevent physicians spending quality time with their patients. He writes,

\begin{quote}
    The increased efficiency and workflow could either be used to squeeze clinicians more, or the gift of time could be turned back to patients — to use the future to bring back the past. The latter objective will require human activism, especially among clinicians, all to stand up for the best interest of patients \citep[21]{topol2019deep}.
\end{quote}

Topol hopes that physicians will mobilize politically to defend their interests – and the interests of patients — in longer conversations about care. We hope so too… but it is vital to acknowledge that this is a hope rather than a prediction. Moreover, there are a number of reasons to believe that is naïve hope.

Political action requires a confident and empowered group of people who share goals and (usually) an identity. Defeats make political action harder, whilst victories make it more likely. Unfortunately, the introduction of AI is likely to demoralise, fragment, and disempower the medical profession at the very point at which Topol expects doctors to rise up and demand better working conditions and outcomes for their patients. One thing that everyone agrees on in discussions of AI in medicine is that its introduction is likely to be highly disruptive to existing practices and institutions \citep[285]{topol2019deep}. Such disruption tends to be unsettling for those who work in the disrupted settings. Even if AI is unlikely to replace physicians entirely \citep{topol2019deep, verghese2018computer}, it is likely to render redundant skills that the current generation of physicians spent years learning and have placed at the heart of their professional self-conception.\footnote{Compare \cite{oecd2016science}, 85, and \cite{cassell2004nature}, 76. See also \cite{wachter2017digital}.} Especially if combined with advances in robotics, AI may also break down complex tasks in healthcare into a number of different tasks that can be performed by people with smaller skill sets, as well as reduce the number of people required to be employed to complete various procedures. More generally, as with previous generations of information and computing technology, the introduction of AI into hospitals and healthcare settings is likely to lead to a shift in power and authority away from frontline practitioners to those who manage and design the IT systems \citep[79-80]{cassell2002doctoring}. Finally, research is already being directed toward using AI to monitor physician performance \citep{dias2019using}, suggesting that physician surveillance will be one of the first uses of AI in the health sector. Physicians who are demoralized, disempowered, concerned for their jobs, and feel themselves to be under surveillance are ill-placed to win political victories.

It must also be observed that the historical record doesn’t inspire much confidence here. Doctors in the US have as yet been unable to motivate US governments to adopt universal basic healthcare or even get the US public to endorse it, despite the fact that universal healthcare would be in the interests of all Americans \citep{lu2003does}. They were unable to resist the rise of managed-care in the 1990s or the destructive impacts of the introduction of electronic medical records in the 2000’s \citep{duijmelinck2016can, friedberg2014factors,hill20134000, verghese2008culture}. Topol himself notes that 

\begin{quote}
    it was, after all, doctors themselves who allowed the invasion of grossly inadequate electronic health records into the clinic, never standing up to companies like Epic, which has, in its contracts with hospitals and doctors, a gag clause that prohibits them from disparaging electronic health records or even publishing EHR screenshots \citep[288]{topol2019deep}.
\end{quote}

This history of failure provides little grounds for confidence that the medical profession will be able to resist the same economic, political, and institutional dynamics when it comes to the adoption of AI. Conversely, if one is concerned about care in medicine, there is little need to await the coming of AI to begin campaigning to defend it.

There are also a number of reasons to think that AI may reduce rather than increase the amount of time that healthcare practitioners have to spend talking with patients.

Most obviously, the fact that the lifeblood of AI is big data suggests that, as AI is introduced, the demand for things to be measured and recorded in medical settings will only increase \citep{maddox2019questions}. That is to say, healthcare workers may be expected to spend more time rather than less staring at screens and filling in forms on computers when they would rather be talking to patients. Again, the lesson of previous generations of technological change, which for the most part have shifted – or even increased – administrative burdens rather than relieved them is relevant here. For instance, when the introduction of computers and electronic health records into hospitals made it easier to record data, the result was that more data was demanded rather than that the same amount of data was recorded more swiftly. Importantly, the operations of AI are themselves likely to generate even more data both about the internal functionings of the systems and about their performance. 

As we observed at the outset, Topol’s hope is that AIs will record and manage all this data by themselves and thus not further burden healthcare workers. There may be some settings where this is the case. However, any optimism here should be tempered by the recognition that one of the lessons of AI research over the last six decades is that “sensing” turns out to be a much harder problem than calculating, planning, or analyzing. Despite remarkable progress in natural language processing in recent years, extracting the meaning of interactions with patients in the clinic in real world conditions, which may require taking into account both patient and physician’s accent, colloquialisms, body language, and social context, remains a formidable challenge. While patients may report to their healthcare providers with more and more data generated by their online-behavior, by apps, and by wearables, working out which datasets are relevant and integrating them with the patient’s medical records often requires human judgement. Until we are prepared to rely entirely on AI for diagnosis, every new scan or test will demand that a clinician looks over the results. In the short-to-medium term, then, AI is likely to require human beings to provide the data that it needs to function.

It’s also possible that the use of AI to gather and record data in some contexts may itself work to the detriment of care. Sometimes physicians gather information by examining, or asking questions of, the patient and this process is also an opportunity for the conversation to roam more widely and thus an opportunity for “care”. This is especially the case where the process of talking to the patient is part of the process of diagnosis or the physical exams that support diagnosis. Gathering this sort of information automatically would actually reduce the opportunities for patients to feel that their physician was genuinely concerned for them \citep{truog2019slide}.

Finally, there is a profound tension in the idea that introducing more machines into the medical setting will lead to better relationships between physicians and patients. This is because AI will tend to undermine trust in doctors and because there are connections, both conceptual and empirical, between care and trust. Notoriously, AIs often function as “black boxes”, with users – and sometimes even their designers – being unable to understand or explain why the AI produces the output that it does. If doctors start to rely on advice from AI the question will arise whether we should — indeed, how we could — trust our doctors. As Watson and colleagues note, “If doctors do not understand why the algorithm made a diagnosis, then why should patients trust the recommended course of treatment?” \citep{watson2019clinical}. If we don’t believe that it is our physician who is really making the decisions about our healthcare, it’s hard to see how we could feel that they are caring for us. They might care about us but that’s not the same as caring for us. 

Indeed, this erosion of trust, with its detrimental impact on care, is likely to happen even if doctors could — if they tried hard enough — explain the outputs of the AI but in practice don’t make the effort to do so. There is a connection here to the question of the likely impact of the introduction of AI on the workload of doctors. If physicians want to retain the trust of their patients and remain the ultimate authority on treatment decisions they will need to supervise and review the operations of AI \citep{verghese2018computer}. At the very least, they will need to be able to assess when the AI is operating properly, which in turn will require being able to access the data on which the AI is relying and check that the conclusions of the AI are plausible in the light of that data. However, the more doctors are expected to do this, the more AI will add to their burden and take their attention away from the patient in front of them \citep{maddox2019questions}. Alternatively, doctors could take the results of the prognostications of AI on faith in the same way they do existing algorithms used in medicines or the conclusions of the peer-reviewed literature. But while patients are used to doctors relying, as we all do, on other people, doctors’ reliance on AI is likely to be more disconcerting, especially as AI comes to take over roles, such as diagnosis, that have traditionally thought to be central to the profession of the physician \citep{wachter2017digital}. If I come to see my doctor as the handmaiden to an AI, which is actually deciding on my treatment, then it may be difficult for me to understand my doctor as providing care.

None of this is to deny the potential of AI to promote any number of other goods in medicine, including, most importantly more timely and accurate diagnosis of a wide range of conditions. Advances in these areas are to be welcomed. Nevertheless, we should be conscious that they may come at a cost to care, given the current pressures on physicians and healthcare providers.

Topol hopes that AI will be used to expand opportunities for care but wishing for something does not make it so. The factors that have led to the decline in human contact in medicine are economic — which is to say, ultimately political — and it is naïve to think that technological change alone is likely to reverse this. If we want to ensure that AI increases the opportunities for, rather than erodes, care in medicine we will need to think deeper, not about AI but about the business of medicine and the institutional and economic contexts in which it is practised today.

\bibliography{main}

\begin{thebibliography}{23}
\providecommand{\natexlab}[1]{#1}
\providecommand{\doi}[1]{\url{https://doi.org/#1}}
\providecommand{\url}[1]{\texttt{#1}}
\providecommand{\urlprefix}{}

\bibitem[{Blythe and Curlin(2018)Blythe, Jacob A and Curlin, Farr A}]{blythe2018just}
Blythe JA, Curlin FA.
\newblock “Just do your job”: Technology, bureaucracy, and the eclipse of conscience in contemporary medicine.
\newblock Theoretical Medicine and Bioethics. 2018;39:431--452.

\bibitem[{Broussard(2018)Broussard, Meredith}]{broussard2018artificial}
Broussard M.
\newblock Artificial unintelligence: How computers misunderstand the world.
\newblock MIT Press; 2018.

\bibitem[{Cassell(2002)Cassell, Eric J}]{cassell2002doctoring}
Cassell EJ.
\newblock Doctoring: The nature of primary care medicine.
\newblock Oxford University Press; 2002.

\bibitem[{Cassell(2004)Cassell, Eric J}]{cassell2004nature}
Cassell EJ.
\newblock The nature of suffering: And the goals of medicine.
\newblock Oxford University Press; 2004.

\bibitem[{Darzi et~al.(2018)Darzi, Ara and Quilter-Pinner, Harry and Kibasi, Tom}]{darzi2018better}
Darzi A, Quilter-Pinner H, Kibasi T.
\newblock Better Health and Care for All: A 10-Point Plan for the 2020s. The Final Report of the Lord Darzi Review of Health and Care.
\newblock Institute of Public Policy Research; 2018.

\bibitem[{Dias et~al.(2019)Dias, Roger D and Gupta, Avni and Yule, Steven J}]{dias2019using}
Dias RD, Gupta A, Yule SJ.
\newblock Using machine learning to assess physician competence: A systematic review.
\newblock Academic Medicine. 2019;94(3):427--439.

\bibitem[{Diefenbach(2009)Diefenbach, Thomas}]{diefenbach2009new}
Diefenbach T.
\newblock New public management in public sector organizations: The dark sides of managerialistic "enlightenment".
\newblock Public Administration. 2009;87(4):892--909.

\bibitem[{Duijmelinck and van~de Ven(2016)Duijmelinck, Dani{\"e}lle and van de Ven, Wynand}]{duijmelinck2016can}
Duijmelinck D, van~de Ven W.
\newblock What can Europe learn from the managed care backlash in the United States?
\newblock Health Policy. 2016;120(5):509--518.

\bibitem[{Friedberg et~al.(2014)Friedberg, Mark W and Chen, Peggy G and Van Busum, Kristin R and Aunon, Frances and Pham, Chau and Caloyeras, John and Mattke, Soeren and Pitchforth, Emma and Quigley, Denise D and Brook, Robert H and others}]{friedberg2014factors}
Friedberg MW, Chen PG, Van~Busum KR, Aunon F, Pham C, Caloyeras J, et~al.
\newblock Factors affecting physician professional satisfaction and their implications for patient care, health systems, and health policy.
\newblock RAND Health Quarterly. 2014;3(4):1.

\bibitem[{Hill~Jr et~al.(2013)Hill Jr, Robert G and Sears, Lynn Marie and Melanson, Scott W}]{hill20134000}
Hill~Jr RG, Sears LM, Melanson SW.
\newblock 4000 clicks: A productivity analysis of electronic medical records in a community hospital ED.
\newblock The American Journal of Emergency Medicine. 2013;31(11):1591--1594.

\bibitem[{Israni and Verghese(2019)Israni, Sonoo Thadaney and Verghese, Abraham}]{israni2019humanizing}
Israni ST, Verghese A.
\newblock Humanizing artificial intelligence.
\newblock JAMA. 2019;321(1):29--30.

\bibitem[{Liu et~al.(2018)Liu, Xiaoxuan and Keane, Pearse A and Denniston, Alastair K}]{liu2018time}
Liu X, Keane PA, Denniston AK.
\newblock Time to regenerate: The doctor in the age of artificial intelligence.
\newblock Journal of the Royal Society of Medicine. 2018;111(4):113--116.

\bibitem[{Lu and Hsiao(2003)Lu, Jui-Fen Rachel and Hsiao, William C}]{lu2003does}
Lu JFR, Hsiao WC.
\newblock Does universal health insurance make health care unaffordable? Lessons from Taiwan.
\newblock Health Affairs. 2003;22(3):77--88.

\bibitem[{Maddox et~al.(2019)Maddox, Thomas M and Rumsfeld, John S and Payne, Philip RO}]{maddox2019questions}
Maddox TM, Rumsfeld JS, Payne PR.
\newblock Questions for artificial intelligence in health care.
\newblock JAMA. 2019;321(1):31--32.

\bibitem[{McKinlay and Marceau(2002)McKinlay, John B and Marceau, Lisa D}]{mckinlay2002end}
McKinlay JB, Marceau LD.
\newblock The end of the golden age of doctoring.
\newblock International Journal of Health Services. 2002;32(2):379--416.

\bibitem[{Mesk{\'o} et~al.(2018)Mesk{\'o}, Bertalan and Het{\'e}nyi, Gergely and Gy{\H{o}}rffy, Zsuzsanna}]{mesko2018will}
Mesk{\'o} B, Het{\'e}nyi G, Gy{\H{o}}rffy Z.
\newblock Will artificial intelligence solve the human resource crisis in healthcare?
\newblock BMC Health Services Research. 2018;18:1--4.

\bibitem[{{Organisation for Economic Co-operation and Development}(2016)}]{oecd2016science}
{Organisation for Economic Co-operation and Development}.
\newblock OECD Science, Technology, and Innovation Outlook 2016.
\newblock Organisation for Economic Co-operation and Development; 2016.

\bibitem[{Topol(2019)Topol, Eric}]{topol2019deep}
Topol E.
\newblock Deep medicine: How artificial intelligence can make healthcare human again.
\newblock Basic Books; 2019.

\bibitem[{Truog(2019)Truog, Robert D}]{truog2019slide}
Truog RD.
\newblock Of slide rules and stethoscopes: AI and the future of doctoring.
\newblock Hastings Center Report. 2019;49(5):3.

\bibitem[{Verghese(2008)Verghese, Abraham}]{verghese2008culture}
Verghese A.
\newblock Culture shock -- Patient as icon, icon as patient.
\newblock The New England Journal of Medicine. 2008;359(26):2748.

\bibitem[{Verghese et~al.(2018)Verghese, Abraham and Shah, Nigam H and Harrington, Robert A}]{verghese2018computer}
Verghese A, Shah NH, Harrington RA.
\newblock What this computer needs is a physician: Humanism and artificial intelligence.
\newblock JAMA. 2018;319(1):19--20.

\bibitem[{Wachter(2017)Wachter, Robert}]{wachter2017digital}
Wachter R.
\newblock The digital doctor: Hope, hype and harm at dawn of medicine's computer age.
\newblock McGraw-Hill; 2017.

\bibitem[{Watson et~al.(2019)Watson, David S and Krutzinna, Jenny and Bruce, Ian N and Griffiths, Christopher EM and McInnes, Iain B and Barnes, Michael R and Floridi, Luciano}]{watson2019clinical}
Watson DS, Krutzinna J, Bruce IN, Griffiths CE, McInnes IB, Barnes MR, et~al.
\newblock Clinical applications of machine learning algorithms: Beyond the black box.
\newblock BMJ. 2019;364.

\end{thebibliography}
\end{document}